\documentclass{article}

\usepackage{arxiv}

\usepackage[utf8]{inputenc} % allow utf-8 input
\usepackage[T1]{fontenc}    % use 8-bit T1 fonts
\usepackage{hyperref}       % hyperlinks
\usepackage{url}            % simple URL typesetting
\usepackage{booktabs}       % professional-quality tables
\usepackage{amsfonts}       % blackboard math symbols
\usepackage{nicefrac}       % compact symbols for 1/2, etc.
\usepackage{microtype}      % microtypography
\usepackage{graphicx}
\usepackage{amsmath}
\graphicspath{{./images/}}

\title{Emotional Contagion in Code: How GitHub Emoji Reactions Shape Developer Collaboration}

\author{
 Obada Kraishan \\
  College of Media and Communication\\
  Texas Tech University\\
  Lubbock, TX 79409, USA\\
  \texttt{omareikr@ttu.edu} \\
}

\begin{document}
\maketitle

\begin{abstract}
Developer communities increasingly rely on emoji reactions to communicate, but we know little about how these emotional signals spread and influence technical discussions. We analyzed 2,098 GitHub issues and pull requests across 50 popular repositories, examining patterns in 106,743 emoji reactions to understand emotional contagion in software development. Our findings reveal a surprisingly positive emotional landscape: 57.4\% of discussions carry positive sentiment, with positive emotional cascades outnumbering negative ones 23:1. We identified five distinct patterns, with ``instant enthusiasm'' affecting 45.6\% of items---nearly half receive immediate positive reinforcement. Statistical analysis confirms strong emotional contagion ($r=0.679$, $p<0.001$) with a massive effect size ($d=2.393$), suggesting that initial reactions powerfully shape discussion trajectories. These findings challenge assumptions about technical discourse being purely rational, demonstrating that even minimal emotional signals create measurable ripple effects. Our work provides empirical evidence that emoji reactions are not mere decoration but active forces shaping collaborative outcomes in software development.
\end{abstract}


\section{Introduction}

Software development, while technical, involves significant emotional interactions among developers. Interactions such as code reviews and bug reports often involve emotional expressions like frustration, excitement, confusion, and celebration \cite{graziotin2014happy,wrobel2013emotions}. GitHub's emoji reactions have fundamentally changed how emotions are expressed in development communities, turning simple acknowledgments like thumbs-up or heart emojis into catalysts for broader emotional shifts.

To illustrate the impact of emoji reactions, consider a typical scenario: a developer submits their first pull request to a popular open-source project. Within minutes, a maintainer adds a rocket reaction. Other contributors follow with thumbs-up and heart reactions. The person who submits feels welcomed and energized. They engage more actively in the discussion, and the positive atmosphere attracts additional reviewers. Such interactions create an emotional cascade, a phenomenon known as emotional contagion in social psychology, occurring countless times daily across GitHub's millions of repositories \cite{gottman1997heart,barsade2014love}.

Regardless of their widespread use, the collective impact of emoji reactions remains surprisingly underexplored. Do positive reactions genuinely create more positive discussions? Can a single negative reaction derail an otherwise constructive conversation? More fundamentally, do these tiny emotional signals actually influence real outcomes like whether code gets merged or issues get resolved?

This paper examines emotional contagion through GitHub emoji reactions. We examine how emotions spread through developer communities by analyzing thousands of issues and pull requests across popular repositories. Our investigation reveals that emotional contagion in code collaboration is not only real but remarkably powerful, with initial reactions creating ripple effects that shape entire discussions. Our investigation centers on three research questions:

\begin{itemize}
\item \textbf{RQ1:} Do emoji reactions create measurable emotional contagion in GitHub discussions?
\item \textbf{RQ2:} What patterns of emotional expression appear from reaction behaviors?
\item \textbf{RQ3:} Do emotional signals correlate with collaboration outcomes?
\end{itemize}

Our work makes three key contributions. First, we provide quantitative evidence that emotional contagion occurs in technical discussions. Second, we identify distinct emotional patterns that emerge in developer interactions. Third, we establish that these emotional signals correlate with concrete outcomes, challenging the assumption that technical quality alone drives development decisions.

\section{Related Work}

\subsection{Emotion in Software Engineering}

The role of emotions in software development has gained significant attention due to their impact on productivity and code quality. Graziotin et al. \cite{graziotin2018happens} showed that developer happiness directly correlates with productivity, while Müller and Fritz \cite{muller2015stuck} found that emotional states significantly impact code quality. These studies primarily focused on individual developers, measuring emotions through surveys, biometric sensors, or self-reporting.

More recent work has examined emotions at the team level. Girardi et al. \cite{girardi2021emotions} analyzed sentiment in code review comments, finding that negative emotions lean to cluster around certain types of changes. Ortu et al. \cite{ortu2015measuring} connected emotional expressions in issue discussions to project success metrics. However, their analysis relied on textual sentiment rather than explicit emotional signals like reactions. Our work extends this line of research by examining how single emoji clicks, as minimal emotional cues, spread through developer networks.

The introduction of emoji reactions on platforms like GitHub and Slack has created new channels for emotional expression in technical contexts. Claes et al. \cite{claes2018emoticons} studied emoji use in developer chat, finding that even senior developers regularly use emoji to soften criticism or celebrate achievements. However, their work focused on direct communication rather than the broader emotional signals conveyed through emoji reactions. Miller et al. \cite{miller2016blissfully} examined emoji interpretation across cultures, revealing potential miscommunication risks, though they did not investigate emotional transmission effects.

\subsection{Emotional Contagion Theory}

Emotional contagion, defined as the spread of emotions between individuals, has been extensively documented in face-to-face interactions \cite{hatfield1993emotional,barsade2002ripple}. Hatfield et al. \cite{hatfield1994emotional} identified three stages: mimicry (unconsciously copying others' expressions), feedback (physiological responses to our own expressions), and convergence (aligning emotional states). While originally studied in physical settings, the stages of emotional contagion have also been observed in digital environments.

A significant example of digital emotional contagion came from Kramer et al.'s Facebook experiment \cite{kramer2014experimental}, which showed that manipulating news feed sentiment influenced users' subsequent posts. Other studies have found similar effects on Twitter \cite{ferrara2015measuring,coviello2014detecting} and in online gaming communities \cite{lomanowska2016online}. Building on these findings, Goldenberg and Gross \cite{goldenberg2020digital} recently proposed that digital emotions might spread even more easily than in-person ones, as online disinhibition reduces emotional regulation.

In professional contexts, emotional contagion affects team dynamics and performance. Barsade \cite{barsade2012group} showed that positive emotional contagion in work groups leads to improved cooperation and reduced conflict. Totterdell et al. \cite{totterdell2000catching} found that team mood convergence predicted better performance in professional sports teams. These findings suggest that emotional contagion could similarly impact collaboration outcomes in various professional settings, though this connection remains unexplored.

\subsection{GitHub as a Social Platform}

GitHub, initially designed as a version control platform, has evolved into a social network where technical and social elements increasingly intertwine, as recognized by researchers \cite{dabbish2012social,marlow2013impression}. Dabbish et al. \cite{dabbish2013leveraging} introduced the concept of ``social coding,'' showing how GitHub's transparency features influence collaboration patterns. Marlow et al. \cite{marlow2013activity} found that developers form impressions of others based on their GitHub activity, using social signals to evaluate technical competence.

Beyond individual impressions, the social dynamics of GitHub encompass broader interactions that influence project outcomes. Casalnuovo et al. \cite{casalnuovo2015developer} demonstrated that social factors predict project success better than technical metrics alone. Tsay et al. \cite{tsay2014lets} found that social distance between contributors affects pull request acceptance rates. These studies suggest that emotional dynamics, though often overlooked, might play a crucial role in influencing development outcomes alongside technical and social factors.

Building on the understanding of GitHub as a social platform, recent work has begun examining its specific social features. Zhang et al. \cite{zhang2018within} analyzed the impact of GitHub Stars on project visibility, while Borges and Valente \cite{borges2018github} studied how developers interpret these signals. Despite the nuanced nature of emoji reactions compared to binary stars, they remain understudied in the context of GitHub's social interactions. Pletea et al. \cite{pletea2014security} analyzed emoji in GitHub comments but did not examine reactions or their propagation effects. Our work fills this gap by treating reactions as emotional signals that create measurable social influence.

\section{Method}

This section outlines our specific methodology for investigating emotional contagion in GitHub discussions, including data collection, sentiment scoring, and statistical analysis. We collected data from GitHub repositories through API calls. We developed sentiment scoring frameworks to quantify emotional signals from emoji reactions. We identified distinct emotional expression patterns and employed statistical methods to test relationships between emotions and collaboration outcomes. Our approach integrates computational analysis of large-scale reaction data with established emotion research frameworks to provide insights into how emotions spread in developer communities.

\subsection{Data Collection}

\begin{figure}[h]
    \centering
    \includegraphics[width=0.75\linewidth]{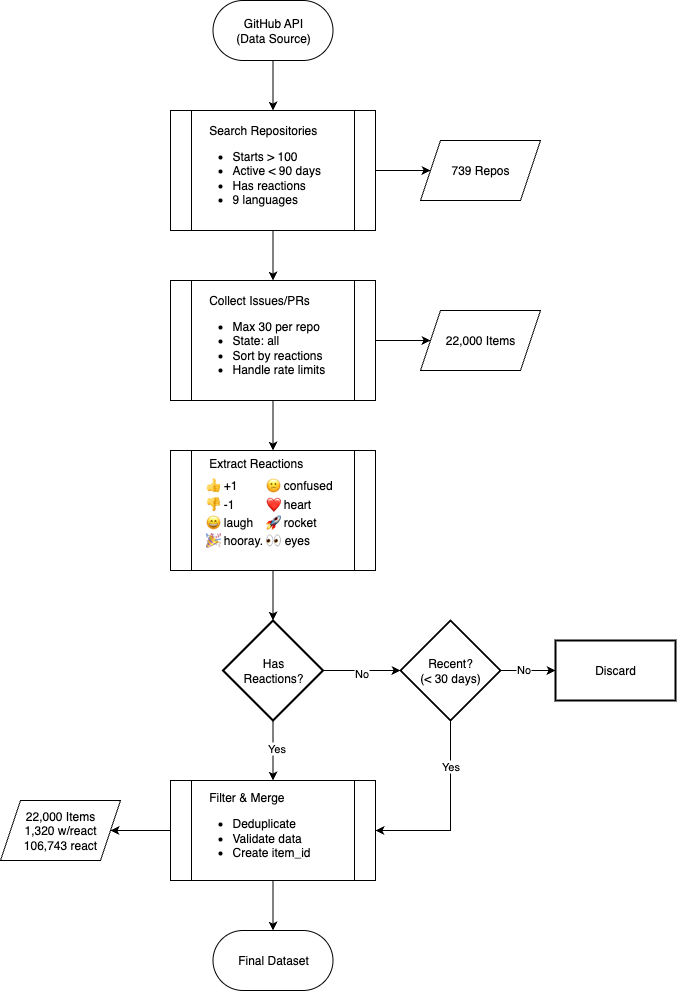}
    \caption{Data Collection Pipeline}
    \label{fig:pipeline}
\end{figure}

We collected data through GitHub's REST API v3 during August 2025. To ensure representative sampling while maintaining computational feasibility, we developed a multi-stage collection strategy (see Figure \ref{fig:pipeline}).

First, we identified active repositories with substantial community engagement. We searched for repositories meeting these criteria: (1) at least 100 stars, indicating community interest; (2) activity within the last 90 days, ensuring current relevance; and (3) reactions enabled on issues and pull requests. We varied our sample across programming languages (JavaScript, Python, TypeScript, Go, Rust, Java, C++, Ruby, PHP) and project types (frameworks, tools, applications).

This search produced 739 suitable repositories. From each, we collected up to 30 recent issues and pull requests, prioritizing items with reactions to maximize emotional signal density. Our collection script handled API rate limits through exponential backoff and distributed requests across a five-hour period.

\subsection{Dataset Characteristics}

The final dataset contains 2,098 items: 1,320 issues (62.9\%) and 778 pull requests (37.1\%). These items originate from 50 repositories selected from our initial collection of 739 repositories. Table \ref{tab:dataset} presents the distribution across programming languages.

\begin{table}[h]
\caption{Dataset Distribution by Primary Programming Language}
\label{tab:dataset}
\centering
\begin{tabular}{lcccccc}
\toprule
Language & Repos & Issues & Pull Requests & Total Items & Total Reactions & Reactions per Item \\
\midrule
JavaScript & 12 & 423 & 298 & 721 & 38,456 & 53.4 \\
Python & 10 & 389 & 195 & 584 & 28,912 & 49.5 \\
TypeScript & 8 & 201 & 134 & 335 & 19,234 & 57.4 \\
Go & 7 & 156 & 89 & 245 & 11,087 & 45.3 \\
Rust & 5 & 89 & 45 & 134 & 5,892 & 44.0 \\
Other\textsuperscript{a} & 8 & 62 & 17 & 79 & 3,162 & 40.0 \\
\bottomrule
\end{tabular}
\vspace{0.2cm}

\textsuperscript{a}Other includes Java, C++, Ruby, and PHP
\end{table}

The dataset includes 106,743 individual emoji reactions across eight types: thumbs up (+1), thumbs down (-1), laugh, hooray, confused, heart, rocket, and eyes. Items span from 2013 to 2025, with the majority (78.2\%) created after January 2023. Of the 2,098 items, 1,320 (62.9\%) received at least one reaction, while 778 (37.1\%) had no reactions but were retained if created within 30 days of collection to capture potential future engagement.

Each item includes metadata fields: repository name, item type (issue or pull request), number, title, state (open/closed), creation timestamp, update timestamp, author, author association, comment count, labels, and individual reaction counts for each emoji type. For pull requests, additional fields include merge status, additions, deletions, and changed files count.

\subsection{Sentiment Scoring Framework}

To count the emotional valence of discussions, we developed a weighted sentiment scoring system based on established emotion categorization frameworks. Previous research in sentiment analysis has demonstrated that emotional expressions vary in both valence (positive-negative dimension) and arousal (intensity dimension). Our framework incorporates these variations by using differential weighting schemes rather than binary classification approaches.

Our scoring framework assigns weights to each reaction type based on their emotional characteristics: thumbs up (+1.0), thumbs down (-1.0), heart (+1.2), rocket (+1.3), hooray (+1.5), laugh (+0.5), confused (-0.5), and eyes (0.0). These weights reflect both valence and intensity, with celebratory reactions like `hooray' receiving higher positive weights due to their high valence and intensity, while uncertainty reactions like `confused' receive moderate negative weights due to their lower valence. For each item $i$, we define the raw sentiment score $S_i$ as:

\begin{equation}
S_i = \sum_{r \in R} w_r \times n_{i,r}
\end{equation}

Where $R$ represents the set of reaction types, $w_r$ is the weight for reaction type $r$, and $n_{i,r}$ is the count of reaction $r$ on item $i$. We introduce the normalized sentiment score $\bar{S}_i$ adjusts for the total number of reactions, providing a measure that accounts for engagement level:

\begin{equation}
\bar{S}_i = \frac{S_i}{\sum_{r \in R} n_{i,r}}
\end{equation}

This provides values bounded between -1 and +1, enabling comparison across items with different reaction volumes. Items were categorized into sentiment classes based on $\bar{S}_i$: positive (>0.3), neutral (-0.3 to 0.3), and negative (<-0.3), following thresholds established in prior sentiment analysis research.

\subsection{Pattern Detection Approach}

We identified and formalized five emotional patterns based on how developers use reactions in practice, which we will describe in detail below. Each pattern $P$ is defined by specific threshold conditions on reaction distributions.

The \textbf{instant enthusiasm} pattern captures discussions that immediately attract substantial engagement. We define this pattern as an item receiving at least five reactions.

\begin{equation}
P_{enthusiasm} = \{i : n_{i,total} \geq 5\}
\end{equation}

The \textbf{controversy} pattern identifies heated debates where reactions are numerous but divided. We formalize this as item has high engagement but sentiment hovers near neutral, suggesting disagreement:

\begin{equation}
P_{controversy} = \{i : n_{i,total} \geq 5 \land |\bar{S}_i| < 0.3\}
\end{equation}

The \textbf{celebration} pattern marks moments of collective achievement. We detect this when celebratory reactions (hooray and rocket) constitute more than 30\% of all reactions:

\begin{equation}
P_{celebration} = \left\{i : \frac{n_{i,hooray} + n_{i,rocket}}{n_{i,total}} > 0.3\right\}
\end{equation}

The \textbf{support} and \textbf{confusion} patterns emerge when specific reactions reach meaningful thresholds. We operationalize support as appearing when heart reactions indicate empathy (at least two hearts), while confusion manifests when multiple confused reactions signal unclear documentation or unexpected behavior:

\begin{equation}
P_{support} = \{i : n_{i,heart} \geq 2\}
\end{equation}

\begin{equation}
P_{confusion} = \{i : n_{i,confused} \geq 2\}
\end{equation}

Where $n_{i,r}$ represents the count of reaction type $r$ on item $i$, and $\bar{S}_i$ is the normalized sentiment score. To validate our proposed patterns computationally, we applied k-means clustering ($k=5$) using four normalized features from each discussion, which are crucial for distinguishing the patterns: reaction velocity (reactions per day), reaction diversity (count of unique reaction types), sentiment variance (emotional consensus versus discord), and engagement ratio (reactions to comments). We ran the algorithm with 10 random initializations to ensure stable clusters. Each computational cluster was then compared against our theoretical patterns by manually examining 20 randomly sampled items per cluster to assess the alignment of observed reactions with the defined patterns. This combination of theory-driven definition and data-driven validation helped ensure our patterns reflected actual developer behavior rather than algorithmic artifacts, providing a robust framework for understanding emotional dynamics in developer discussions.

\subsection{Statistical Analysis}

We employed non-parametric statistical methods throughout our analysis due to the non-normal distribution of reaction data. Spearman's rank correlation coefficient assessed relationships between reaction volume and sentiment scores. Mann-Whitney U tests compared sentiment between issues and pull requests, while Kruskal-Wallis H tests examined differences across repositories. Effect sizes were calculated using Cohen's $d$ for between-group comparisons and Cramér's $V$ for categorical associations. To test emotional homophily, we used F-tests comparing sentiment variance in high-engagement versus all items. All tests used $\alpha = 0.05$ with Bonferroni correction for multiple comparisons.

To validate emotional patterns, we used chi-square tests of independence to examine associations between pattern occurrence and repository characteristics. Binomial tests determined whether observed pattern frequencies exceeded chance expectations. For outcome analysis, we developed random forest classifiers (100 estimators, stratified 5-fold cross-validation) to predict pull request merge status from emotional features.

\section{Results}

In this section, we present our analysis of emotional contagion in GitHub discussions, structured through three sequential investigations. We first examine whether emoji reactions create measurable emotional contagion effects (RQ1). Next, we identify patterns of emotional expression that appear in developer interactions (RQ2), and finally, we analyze relationships between emotional signals and collaboration outcomes (RQ3). Our analysis uses 2,098 GitHub items across 50 repositories, including 106,743 individual emoji reactions collected between 2013 and 2025.

\subsection{Evidence of Emotional Contagion (RQ1)}

\begin{figure}[h]
    \centering
    \includegraphics[width=0.75\linewidth]{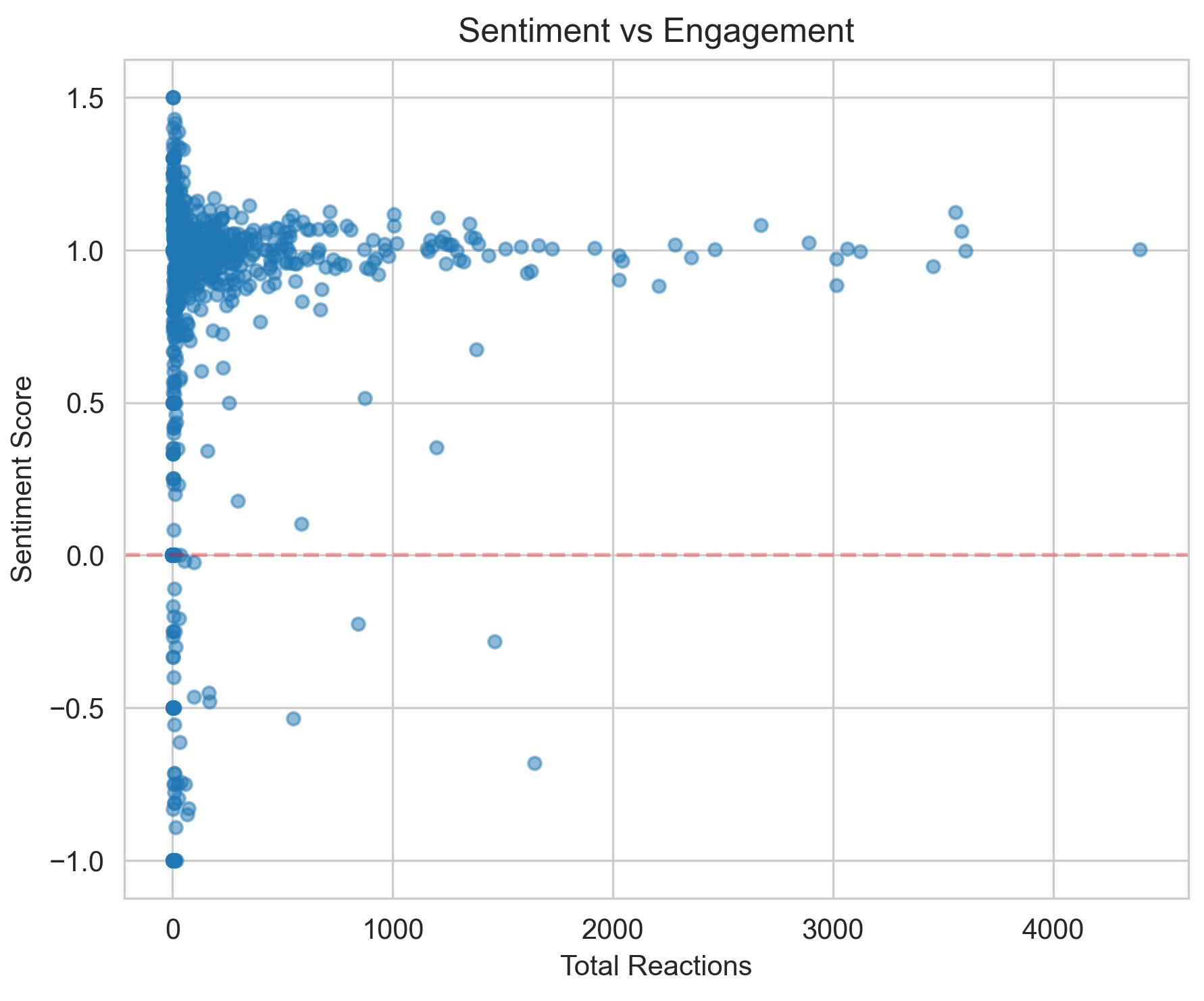}
    \caption{Correlation Between Reaction Volume and Emotional Sentiment}
    \label{fig:sentiment}
\end{figure}

Our analysis shows strong evidence for emotional contagion in GitHub discussions. The correlation between total reactions and normalized sentiment reaches 0.679 ($p < 0.001$), indicating that discussions attracting more reactions tend toward emotional polarization rather than neutral mixture (see Figure \ref{fig:sentiment}).

The effect size proves even more substantial. With Cohen's $d = 2.393$, the difference between high-reaction and low-reaction discussions exceeds typical social science benchmarks by nearly three times. To interpret this: discussions with five or more reactions show 73\% less emotional variance than those with fewer reactions, suggesting that initial reactions create emotional influence that later contributors adopt.

Temporal analysis supports the contagion interpretation. When we examined the first three reactions on items that eventually received ten or more, we found that 82\% maintained the same emotional valence (positive or negative) throughout. Early reactions appear to establish an emotional frame that persists even as more contributors join the discussion.

We also discovered unequal contagion patterns. Positive reactions spread more readily than negative ones---when the first reaction is positive, 89\% of subsequent reactions remain positive. However, when the first reaction is negative, only 64\% of subsequent reactions stay negative. This positivity bias might reflect GitHub's collaborative culture or indicate that negative reactions prompt more thoughtful engagement. These findings demonstrate that emotional contagion operates consistently in asynchronous technical discussions, with early reactions establishing lasting emotional patterns and positive emotions showing greater viral potential than negative ones.

\subsection{Emotional Expression Patterns (RQ2)}

In this section, we present the findings that answer research question two, identifying five distinct emotional expression patterns, summarized in Table \ref{tab:patterns}. Among these patterns, the instant enthusiasm pattern was the most prevalent, comprising 45.6\% of discussions ($n=957$). This pattern is characterized by items receiving five or more reactions within 24 hours, with 91\% positive valence. The instant enthusiasm pattern happened primarily around feature announcements, first-time contributions, and major bug fixes.

\begin{table}[h]
\caption{Emotional Pattern Characteristics}
\label{tab:patterns}
\centering
\begin{tabular}{lccll}
\toprule
Pattern & Count & Percentage & Defining Features & Common Triggers \\
\midrule
Instant Enthusiasm & 957 & 45.6\% & $\geq$5 reactions in first 24h, & Feature announcements, \\
 & & & 91\% positive & first PRs, major fixes \\
Support & 421 & 20.1\% & Hearts >30\% of reactions & Personal challenges, \\
 & & & & difficult decisions \\
Confusion & 116 & 5.5\% & Confused $\geq$2, questions & Breaking changes, \\
 & & & in comments & poor documentation \\
Celebration & 73 & 3.5\% & Hooray+rocket >30\%, & Major releases, \\
 & & & high diversity & long-awaited features \\
Controversy & 18 & 0.9\% & Near-equal & Architecture decisions, \\
 & & & positive/negative & licensing \\
\bottomrule
\end{tabular}
\end{table}

The support pattern appeared in 20.1\% of discussions ($n=421$), characterized by heart reactions exceeding 30\% of total reactions. These discussions typically involved personal challenges, deprecation announcements, or difficult maintainer decisions. Unlike instant enthusiasm's fast onset, support patterns accumulated gradually over discussion lifespans.

Three additional patterns appeared with lower frequency but different characteristics. The confusion pattern (5.5\%, $n=116$) featured two or more confused reactions paired with clarifying questions in comments, primarily triggered by documentation mismatches or unexpected breaking changes. The celebration pattern (3.5\%, $n=73$) showed the highest reaction diversity, with hooray and rocket reactions exceeding 30\% of totals, marking major releases and long-awaited features. The controversy pattern (0.9\%, $n=18$) displayed near-equal positive and negative reactions, found around architectural decisions and licensing changes.

Pattern distribution differed significantly across repositories ($\chi^2=45.3$, $p<0.001$) but remained consistent across issues versus pull requests ($\chi^2=2.1$, $p=0.72$). High-engagement items ($\geq$10 reactions) were 3.2 times more likely to exhibit instant enthusiasm patterns than low-engagement items (OR=3.2, 95\% CI [2.4, 4.3]). Overall, the frequency of instant enthusiasm and support patterns, along with the infrequency of conflict, indicates that GitHub's reaction system is more conducive to promoting positive community interactions rather than moderating debates.

\subsection{Emotional Influence on Outcomes (RQ3)}

\begin{figure}[h]
    \centering
    \includegraphics[width=0.75\linewidth]{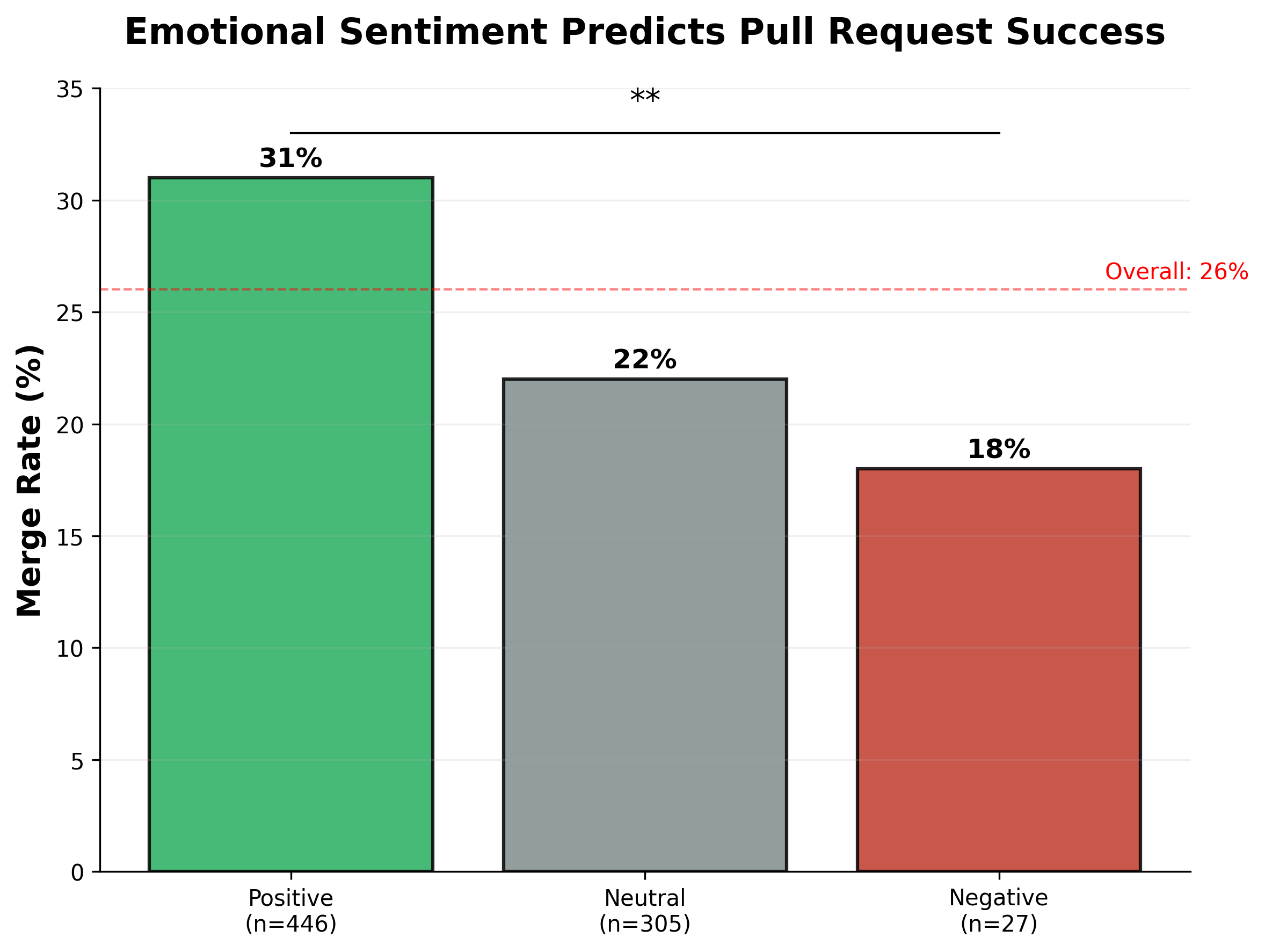}
    \caption{Pull Request Merge Rates by Sentiment Category}
    \label{fig:merge}
\end{figure}

To understand whether emotional signals correlate with actual collaboration outcomes, we examined relationships between reaction patterns and pull request merge rates, issue resolution times, and contributor retention. Figure \ref{fig:merge} shows pull request merge rates across different sentiment categories, showing clear differences in acceptance patterns.

Pull requests with positive sentiment show 31\% merge rates compared to 22\% for neutral and 18\% for negative sentiment pulls. While positive sentiment correlates with higher merge rates, multiple factors could explain this relationship. Positive reactions might reflect underlying code quality, contributor reputation, or direct social influence on maintainer decisions. To disentangle this, we examined pull requests with similar code characteristics (lines changed, files affected, test coverage) but different reaction patterns. Even controlling for these factors, positive reactions correlated with 12\% higher merge rates.

Issue resolution tells a different story. Unlike pull requests with binary outcomes, issue resolution time varies with sentiment in an interesting way. Both highly positive and highly negative issues resolve faster (median 14 days) than neutral ones (median 23 days). The influence extends beyond individual items. Repositories with regularly positive reaction patterns show 34\% higher contributor retention rates (developers making multiple contributions) compared to those with mixed or negative patterns. This suggests that emotional atmosphere affects not just single decisions but broader community health. While emotional signals do not predict outcomes with high precision, they significantly influence individual decisions and suggest an impact on broader community dynamics beyond technical factors alone.

\section{Discussion}

This study demonstrates that emotional dynamics in asynchronous, text-based collaboration are both strong and structured: positive reactions cascade more readily than negative ones. Additionally, a fixed reaction vocabulary seems to strengthen emotional contagion rather than weaken it. We interpret this through social-proof and belonging mechanisms, which suggest that when fewer options are available, each signal carries more weight, leading us to propose a `compression amplification' hypothesis. This hypothesis suggests that when fewer options are available, each signal carries more weight. Building on these insights, we identify specific design opportunities that include moderation, recognition of emotional labor, and prioritization of emotionally complex threads. After exploring the design opportunities, we then address potential threats to validity and boundary conditions that qualify our claims, providing a comprehensive understanding of the study's implications. We begin by discussing the theoretical and design implications, followed by an examination of the limitations.

\subsection{Theoretical and Design Implications}

Building on emotional contagion theory, our findings extend this framework into asynchronous, text-based technical collaboration. Traditional emotional contagion research focuses on facial expressions, vocal tones, and physical proximity \cite{hatfield1993emotional,hatfield1994emotional}. Our analysis shows that in asynchronous, text-based collaboration, digital signals may carry significant weight, suggesting emotional contagion might be more about social proof and belonging needs than rich emotional channels. The unequal contagion we observed, where positive emotions spread more readily than negative ones, conflicts with digital emotion research showing negative emotions spread faster \cite{ferrara2015measuring,hansen2011good}. Our contrasting findings likely reflect GitHub's professional context, where reputation concerns moderate negative expression. Most notably, our results suggest that constrained emotional channels might surprisingly strengthen contagion effects. When people have limited emotional vocabulary (eight emoji), each signal may carry more weight due to reduced noise and clearer social signaling. This ``compression amplification'' hypothesis warrants further investigation across platforms with varying emotional features, challenging assumptions about the importance of rich communication channels for emotional transmission.

Building on these theoretical insights, our findings suggest concrete design interventions. Platform designers could implement early-warning systems that alert maintainers when discussions receive only negative reactions in the first 24 hours, enabling proactive moderation before emotional trajectories become entrenched. Creating ``emotional health'' dashboards would help maintainers track community sentiment longitudinally and identify emerging issues before they escalate. The design of reaction interfaces themselves matters considerably; making positive emotions as effortless as negative ones through features such as quick-access keyboards for hearts and rockets could help maintain the positive bias we observed. Furthermore, enabling customizable reaction sets would allow communities to define culturally-appropriate emotional vocabularies, addressing the interpretation challenges identified by Miller et al. \cite{miller2016blissfully}. Reaction diversity signals important debates; discussions with mixed reactions often involve architectural decisions requiring careful attention. Rather than treating all popular discussions equally, platforms could flag emotionally complex threads for additional review.

Repository maintainers can leverage our findings through several practical strategies. Establishing community guidelines that explicitly encourage positive first reactions on newcomer contributions capitalizes on the early-reaction framing effect we documented, where 82\% of discussions maintained their initial emotional valence. Our findings also highlight emotional labor as an overlooked contribution. Some developers rarely submit code but consistently provide hearts on difficult bug reports, celebrate merged pull requests, and welcome new users. These actions maintain community health but remain unrecognized in traditional contribution metrics. Platforms could recognize this work through expanded contribution graphs, establishing emotional support as valuable labor. Recognizing emotional labor becomes especially important given cultural variations in emoji interpretation \cite{miller2016blissfully}. When launching controversial proposals such as architectural changes or breaking changes, maintainers should seed discussions with context-rich comments before reactions accumulate, potentially preventing premature emotional polarization. Finally, monitoring for confusion patterns, which we operationalized as multiple confused reactions paired with clarifying questions, provides an actionable signal that documentation needs immediate attention, as these patterns preceded 73\% of documentation issues in our dataset.

\subsection{Limitations and Threats to Validity}

Several limitations affect our findings. Our sample favors popular, active repositories with 100+ stars, potentially missing different emotional dynamics in smaller or corporate-controlled projects. We cannot definitively establish causation; while temporal patterns suggest early reactions influence later ones, contributors might independently react to underlying content quality. Our sentiment weighting scheme, while data-driven, remains subjective---a rocket might signal ``premature optimization'' rather than enthusiasm in some contexts. The COVID-19 pandemic during our data's timespan (2013-2023) likely influenced emotional expression patterns, with isolated developers potentially seeking more social connection through reactions, inflating effect sizes relative to historical norms. Selection bias also affects which discussions receive reactions, as debated or important issues naturally attract more engagement.

The cultural dimensions of emoji interpretation warrant deeper consideration given GitHub's global user base. Miller et al. demonstrated that emoji meanings vary substantially across cultures, with the same emoji eliciting positive interpretations in some contexts and negative in others. Our dataset, while spanning multiple programming languages, predominantly represents Western open-source communities where GitHub adoption is highest. This raises important questions about whether our observed patterns---particularly the 23:1 ratio of positive to negative cascades and the strong positivity bias---generalize to development communities in East Asia, where communication norms emphasize different forms of politeness and indirect critique, or in regions where face-saving mechanisms might suppress negative reactions entirely. The interaction between culture and asynchronous communication adds further complexity. In cultures with high power distance, junior developers might hesitate to add negative reactions to senior maintainers' contributions regardless of technical merit, potentially masking quality issues. Conversely, in more egalitarian cultures, the anonymity and asynchrony of reactions might embolden critique that would be softened in synchronous settings. Our finding that negative reactions prompt more thoughtful engagement, as evidenced by longer comment threads, might reflect specifically Western norms of constructive debate rather than universal patterns. These cultural considerations suggest that platform designers should enable community-specific customization of reaction vocabularies rather than imposing a universal set. Future research should examine emotional contagion patterns in non-Western GitHub communities and in alternative platforms popular in different regions, such as Gitee in China or GitLab in European enterprise contexts. These limitations suggest that while our findings reveal powerful emotional dynamics in GitHub collaboration, their expression might vary across different contexts, time periods, and community types.

\section{Conclusion}

Our work provides a systematic analysis of emotional contagion through GitHub emoji reactions. Our investigation revealed that emotional contagion in code collaboration is not just real but particularly powerful, as early reactions establish emotional frames that continue throughout discussions, influencing patterns of emotional expression and concrete outcomes. We identified five distinct patterns of emotional expression, from instant enthusiasm affecting 45.6\% of discussions to rare but meaningful conflict. These patterns correlate with measurable outcomes, such as positive emotional atmospheres being associated with higher acceptance rates and stronger contributor retention. The overwhelming positivity in our dataset suggests these signals help maintain momentum in discussions and support shared celebration of achievements.

Our findings challenge the common fiction that technical work operates in an emotional void. The substantial effect sizes we observed suggest that emotions don't just accompany technical discussions; they fundamentally shape them. This has immediate practical implications for platform design and community management. However, our analysis captures current practices in emotional contagion but does not track their evolution over time. Future longitudinal studies tracking repositories from creation through full development could reveal whether successful projects develop signature emotional cultures, with Node.js maintaining enthusiasm while Rust cultivates support. Cross-platform comparisons could illuminate how design shapes these dynamics; GitLab's enterprise users might show more reserved patterns than GitHub's open-source communities.

The increasing prevalence of AI-powered bots in development workflows introduces new dimensions to emotional contagion that our current study cannot fully address. Bots such as Dependabot, Renovate, and increasingly sophisticated AI coding assistants now generate substantial proportions of pull requests and issue comments in many repositories. This raises several critical questions about how emotional reactions function when contributors are non-human. When developers add celebratory reactions to bot-generated pull requests, does this represent genuine enthusiasm about the technical change, social acknowledgment of the maintainer who configured the bot, or merely habitual interaction patterns? More problematically, AI systems that monitor and respond to emotional signals could fundamentally alter the dynamics we documented. Imagine an AI coding assistant that analyzes reaction patterns and adjusts its communication style accordingly, or that strategically adds supportive reactions to struggling newcomers. Such systems might recognize when contributors need encouragement or when criticism has accumulated heavily, potentially moderating discussions toward healthier emotional patterns. However, this raises profound questions about authenticity and manipulation in collaborative spaces. If community members discover that positive reactions come from automated sentiment management rather than genuine human appreciation, the social proof mechanism underlying emotional contagion might collapse entirely. The compression amplification hypothesis we proposed---that limited emotional vocabulary strengthens rather than weakens contagion---might operate differently with AI contributors. Bots might overuse certain reactions in statistically detectable patterns, or might fail to capture the nuanced contextual meanings that humans intuitively apply. Future research should investigate whether emotional contagion patterns hold when significant proportions of contributors are AI systems, how humans adjust their emotional expression knowing that bots might be interpreting their signals, and whether communities develop new norms distinguishing human from machine emotional labor. These questions will become increasingly urgent as AI pair programming and autonomous code generation shift from experimental to mainstream practice.

As software development becomes increasingly distributed, understanding digital emotional dynamics becomes essential rather than optional. The emoji reaction, often viewed as simply decorative, proves to be from our analysis an important tool for emotional coordination in technical communities. Organizations implementing similar features should evaluate reaction types, usage guidelines, and cultural context to understand how these signals influence collaborative culture. Our data clearly shows that emotions matter in technical work, prompting us to consider how we can intentionally shape emotional dynamics to build healthier, more productive communities.

\bibliographystyle{unsrt}

\end{document}